\documentclass[
tightenlines,
nofootinbib,
twocolumn,
 amsmath,amssymb,
 aps,
]{revtex4-1}

\usepackage{amsmath}
\usepackage{amssymb}
\usepackage{bm}
\usepackage{color,graphicx}
\usepackage{colortbl}
\usepackage{slashed}
\usepackage{comment}
\usepackage{mathtools}
\usepackage{booktabs}

\newcounter{comment}

\begin{document}

\title{Deep Learning Analysis of Deeply Virtual Exclusive Photoproduction}

\author{Jake Grigsby}
\email{jcg6dn@virginia.edu}
\affiliation{Department of Physics, University of Virginia, Charlottesville, VA 22904, USA.}

\author{Brandon Kriesten} 
\email{btk8bh@virginia.edu}
\affiliation{Department of Physics, University of Virginia, Charlottesville, VA 22904, USA.}

\author{Peter Alonzi} 
\email{lpa2a@virginia.edu}
\affiliation{School of Data Science, University of Virginia, Charlottesville, VA 22904, USA}

\author{Matthias Burkardt} 
\email{burkardt@nmsu.edu}
\affiliation{New Mexico State University - Department of Physics, Box 30001 MSC 3D, Las Cruces NM, 88003 - USA}

\author{Joshua Hoskins} 
\email{jhoskins@jlab.org}
\affiliation{Department of Physics, University of Virginia, Charlottesville, VA 22904, USA.}

\author{Simonetta Liuti} 
\email{sl4y@virginia.edu}
\affiliation{Department of Physics, University of Virginia, Charlottesville, VA 22904, USA.}


\begin{abstract}
We present a Machine Learning based approach to the cross section and asymmetries for deeply virtual Compton scattering from an unpolarized proton target using both an unpolarized and polarized electron beam. 
Machine learning methods are needed to study and eventually interpret the outcome of deeply virtual exclusive experiments since
these reactions are characterized by a complex final state with a larger number of kinematic variables and observables, exponentially increasing the difficulty of quantitative analyses.
Our deep neural network (FemtoNet) uncovers emergent features in the data and learns an accurate approximation of the cross section that outperforms standard baselines. 
%
FemtoNet reveals that the predictions in the unpolarized case systematically show a smaller relative median error than the polarized that can be ascribed to the presence of the Bethe Heitler process. It also suggests that the $t$ dependence can be more easily extrapolated than for the other variables, namely the skewness, $\xi$ and four-momentum transfer, $Q^2$.
%
Our approach is fully scalable and will be capable of handling larger data sets as they are released from future experiments.
 \end{abstract}

\maketitle

\allowdisplaybreaks

\section{Introduction}
Artificial Intelligence (AI) methods including  machine learning (ML), deep learning (DL) and neural networks (NN) have been increasingly applied to 
attack several of the outstanding questions on  
strongly interacting systems in Quantum Chromodynamics (QCD),  ranging from both the sign problem (see \cite{Ohnishi:2019ljc,Mori:2017pne,Alexandru:2018fqp} and references therein), and the inverse problem in lattice QCD, to extracting the behavior of parton distributions functions (PDFs) from data. 
%
%
However, the systematic application of AI to problems in hadronic physics can still be considered to be in its initial stages, which involves exploring a variety of methodologies \cite{Bedaque:2020pct}. One important aspect that is still being developed concerns the treatment of uncertainty, in particular, studying the accuracy of AI based predictions within a well defined statistical approach and with precise benchmarking.



In this paper we address the recently developed science of nuclear femtography, which aims at reconstructing the spatial structure of hadrons, or nucleon tomography, using information from deeply virtual electron proton exclusive photoproduction, $ e p \rightarrow e' p' \gamma$, and various other related reactions. The open questions that we address concern, in particular, the choice of a ML method that is suitable to attack the issues specific to nuclear femtography, including the impact of noise and experimental uncertainty.

Deeply virtual exclusive processes have been identified as the cleanest experimental probe of the 3D structure of the proton as well as its mechanical properties including the quark and gluon distributions of angular momentum, pressure, and shear forces (see reviews in \cite{Goeke:2001tz,Diehl:2001pm,Belitsky:2005qn,Kumericki:2016ehc}). Information on these distributions is encoded in the Generalized Parton Distributions (GPDs) which are functions of the longitudinal momentum fraction, $x$, and of the longitudinal momentum transfer, $\xi$ (skewness parameter) and four momentum transfer squared, $t=\Delta^2$, at a given scale determined by the photon virtuality, $Q^2$. GPDs can be defined for all the different polarization configurations of the struck quark/gluon and proton. Their Fourier transform with respect to the transverse component of $\Delta$ produces faithful images of the proton constituents structure on the transverse plane. 

In exclusive photoproduction we access GPDs from the matrix elements for deeply virtual Compton scattering (DVCS), where the photon is produced at the proton vertex. A competing background process given by the Bethe-Heitler (BH) reaction is also present, where the photon is emitted from the electron and the matrix elements measure the proton elastic form factors. Note that for this prototypical process, the cross section is a function of four independent kinematic variables: the angle between the lepton and hadron planes, $\phi$, the scale, $Q^2$, the skewness, $\xi$, which is related to Bjorken $x$, and $t$; GPDs for all quark flavors appear in the cross section embedded in convolutions over $x$ known as Compton form factors (CFFs), with complex kernels depending on ($x, \xi, Q^2$). Several CFF combinations corresponding to various quark-proton polarization configurations appear simultaneously in the cross section terms for each beam-target polarization configuration \cite{Kriesten:2019jep}. The complexity associated with the large number of kinematic variables, and the complex structure of the cross section makes the problem of extracting tomographic images of quark and gluon structure of the proton virtually impossible to solve with traditional methods; for high precision femtography, which is required to obtain proton images, we need to develop more sophisticated analyses. The success of ML methodologies in modeling complex natural phenomenon make this a prime choice for GPD extraction.

The ultimate goal is development of a tool set for the extraction of GPD from experimental data however, as a first step, we develop a framework for the extraction of the scattering cross-section. The results of our analysis show that the network can effectively generalize in $t$, even in the regions with no data, but is less effective in $\xi$, or $Q^2$. These results are not only beneficial to the physics community but provide interesting overlays with the data science community. One important example is discussion of handling experimental error bars found in Section \ref{sec:baselines}, which is ubiquitous to physics analyses but less commonly considered in building machine learning models.


We use industry standard ML techniques to fit a cross section model based on currently available DVCS experimental data, allowing us to make efficient and accurate predictions across a wide kinematic range represented by the data. By estimating model uncertainty, we can make informed decisions about predictions well outside of the region defined by our data. Several of the practical challenges of fitting a sparse Neural Network, with significant experimental uncertainty, are addressed in this paper. All of our software tools are open-source and can be found \cite{femtonet}.

This paper is organized as follows: in Section \ref{sec:2} we present the theoretical framework for DVCS and give background on Deep Learning and Neural Networks. In Section \ref{sec:3} we detail the specifics of our cross section network and training process. Then, in Section \ref{sec:4} we use our networks to make predictions across a wide range of kinematic bins, comparing them against both experimental measurements and previous theory-driven models. Finally, our conclusions and outlook are presented in Section \ref{sec:5}.

\section{Background}
\label{sec:2}
\subsection{DVCS Theoretical Framework}
The cross section for $ep \rightarrow e'  p' \gamma$ reads,

\begin{equation}
\label{eq:xs}
\frac{d^5\sigma}{d x_{Bj} d Q^2 d|t| d\varphi d \varphi_S } =
\Gamma \,  
\big|T\big|^2 \;.
\end{equation}
where $T$ is a superposition of the DVCS amplitude and it's background process (Bethe-Heitler) BH,
\begin{equation}
|T|^2 = |T_{\rm BH} + T_{\rm DVCS}|^2
=|T_{\rm BH}|^2 + |T_{\rm DVCS}|^2 + \mathcal{I}\;
\label{eq:xsx}
\end{equation}
\begin{eqnarray}
\mathcal{I} & = & T_{BH}^{*} T_{DVCS}
+ T_{DVCS}^{*} T_{BH} 
\end{eqnarray}
The cross section is a differential form of five independent variables, the four momentum transfer squared between the initial ($e$) and final ($e'$) electrons, $Q^2$; $x_{Bj}= Q^2/(2M\nu)$ where $\nu$ is the energy transfer between the initial and final electrons and $M$ is the proton mass; $t$, the (negative) four momentum transfer squared between the initial and final proton; the angle $\varphi$ between the lepton and hadron planes, and $\varphi_S$, the azimuthal angle of the transverse proton spin vector. We will consider in our analysis available fixed target data for an unpolarized target where the measured set of observables includes 
\[E_e, Q^2, x_{Bj}, t, \varphi,\] 

\noindent $E_e$  being the initial electron energy in the lab (the angle $\varphi_S$ is integrated over since we are not considering scattering from a polarized target). The ultimate physics goal is to extract from experiment the QCD CFFs which appear in the DVCS matrix element. At leading order one can define eight CFFs, which correspond to the real and imaginary parts of the two vector and two axial vector current components, respectively. 
Before considering the full extraction of all eight CFFs, an important first step is to understand what is required to learn the form of the cross section. 
Our ML analysis yields model independent results for the cross section independent from the separation into the BH, DVCS and ${\cal I}$ components given in Eq.(\ref{eq:xsx}). 
Because of the nature of the scattering experiments there are many more data on the $\phi$ dependence than on the other variables (see Table I in Section \ref{feature-engineering}). 
In this paper we focus on the simplest case of unpolarized target scattering for which we also have the largest DVCS dataset.

\subsection{Deep Neural Networks}
\label{sec:background}

A Deep Neural Network (DNN) is an effective function approximation method consisting of several layers of computation, in which each layer computes a new representation of the previous layer's outputs. In the simplest case, each layer consists of an affine transformation of the previous layer's representation of the data, followed by an element-wise nonlinearity. For example, the output of the first layer is:
\begin{align}
    \mathbf{x}_1 = z(\mathbf{W}_0\mathbf{x} + \mathbf{b}_0)
\end{align}
Where $z$ is a nonlinear activation function applied to each element of its input vector. $\mathbf{W}$ is an $m \times n$ matrix where $m$ is the dimension of the previous layer's representation and $n$ is the dimension of the current layer's output. $\mathbf{b}$ is a vector of size $n$. The entire model can be written as a composition of functions
\begin{eqnarray}
\label{ff}
   f_{\theta}(\mathbf{x}) = &\mathbf{W}_k(z(\mathbf{W}_{k-1}(\dots(z(\mathbf{W}_0\mathbf{x}
    &+ \mathbf{b}_0))) + \mathbf{b}_{k-1})) + \mathbf{b}_k , \nonumber \\
\end{eqnarray}

\noindent
where $k$ is the number of hidden layers in the model. Model `architectures' are descriptions of the number ($k$) and size ($n$) of a network's layers, as well as the choice of nonlinearity ($z$). $\theta$ is the set of all the entries in the transformation weights $\mathbf{W}$ and biases $\mathbf{b}$. These parameters are initialized randomly, and then adjusted by an optimization algorithm. The goal is to minimize some metric of prediction error - usually called the loss function. A standard choice for regression problems is the Mean Squared Error (MSE):

\begin{eqnarray}
\mathcal{L}_{\theta}^{MSE}(\textbf{x},y) = \frac{1}{2}(y - f_{\theta}(\textbf{x}))^{2}
\end{eqnarray}

Where $y$ is the true value of our predicted variable, often called the `label' or `target' value. In our case, backpropagation is used to compute the partial derivatives of the loss function on a batch of $B$ samples w.r.t each of the parameters in $\theta$, and the resulting gradient vector is used to update the parameters via Stochastic Gradient Descent (SGD):

\begin{eqnarray}
\theta_{t+1} = \theta_t - \lambda \nabla_{\theta} \frac{1}{B} \displaystyle \Sigma_{i=0}^{B}{\mathcal{L}_{\theta_t}(\textbf{x}^{(i)}, \sigma^i)}
\end{eqnarray}

Where $\lambda$ is a learning rate parameter. Modern optimizers \citep{kingma2014adam} \citep{hinton2012neural} often dynamically adjust $\lambda$ based on heuristics like momentum, where consecutive gradient steps in a similar direction increase the learning rate and rapid changes in direction push it towards zero. 

 


\section{Training the Deep Network}
\label{sec:3}

\subsection{Experimental Data}
\label{feature-engineering}
\begin{figure}[h!]
    \centering
    \includegraphics[scale = 0.38]{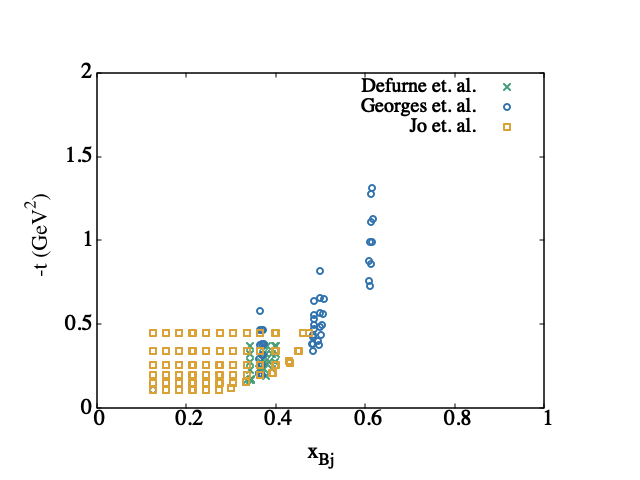}
    \includegraphics[scale = 0.38]{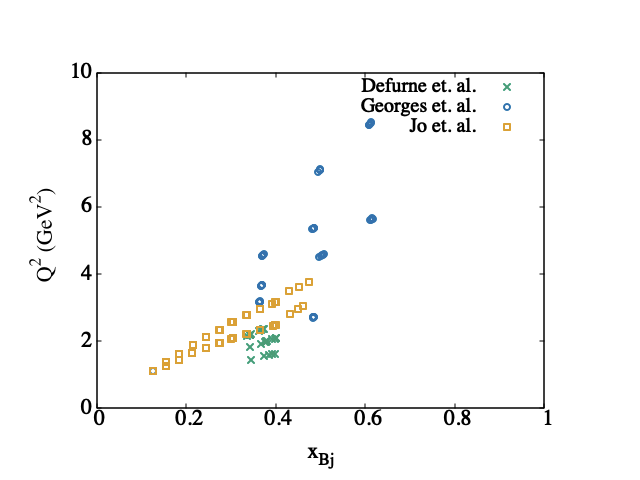}
    \caption{({\it{Top}}) Kinematic region in $x_{Bj}$ and $t$ and  ({\it{Bottom}}) $x_{Bj}$ and $Q^{2}$ where experimental measurements used in this analysis have been taken at 6 GeV and 12 GeV.}
    \label{fig:kin_plots}
\end{figure}

We compiled a dataset from several DVCS experiments \cite{Georges:2018kyi, Defurne:2015kxq, Defurne:2017paw, Jo:2012yt}, covering the range of kinematic bins shown in Figure \ref{fig:kin_plots}. We discarded data sets that listed no explicit $\varphi$ dependence. We used a common format for collecting and recording our data sets such that they list the kinematic dependence of the cross section points, the phi dependence, the polarization of the observable, and the statistical and systematic errors. 

\begin{widetext}
\begin{center}
\begin{scriptsize}
\centering
\begin{table*}[ht]
\begin{tabular}{|c|c|c|c|c|c|c|c|}
\hline
\hline
Experiment          & $Q^{2}$ (GeV$^{2}$) & $t_{\text{min}} - t_{\text{max}}$ (GeV$^{2}$) & $x_{\text{Bj}}^{\text{min}} - x_{\text{Bj}}^{\text{max}}$ & $E_{\text{beam}}$ (GeV)          & Polarization & \# points & Reference \\ \hline \hline
Hall B         & 1.11 / 3.77         & -0.11 / -0.45                                 & 0.13 / 0.48                                   & 5.75                             & UU / LU      & 3862     &    \cite{Jo:2012yt}       \\ \hline
Hall A  & 1.45 / 2.38         & -0.17 / -0.37                                 & 0.34 / 0.40                                   & 5.75                             & UU / LU      & 936      &   \cite{Defurne:2015kxq}        \\ \hline
Hall A  & 1.51 / 2.00         & -0.18 / -0.36                                 & 0.36                                          & 5.55                             & UU / LU      & 788      &      \cite{Defurne:2017paw}     \\ \hline
Hall A    & 2.71 / 8.51         & -0.21 / -1.28                                 & 0.34 / 0.61                                   & 4.49 / 10.99 & UU / LU      & 2160     &  \cite{Georges:2018kyi}         \\ \hline \hline
\end{tabular}
\label{table:data}
\caption{Experimental data used in our neural network analysis. No data from HERMES or COMPAS was used because it is integrated in $\varphi$ and therefore inconsistent dimensionally with a majority of the JLAB DVCS data.}
\end{table*}
\end{scriptsize}
\end{center}
\end{widetext}

 Each entry consists of the input kinematic variables ($x_{Bj}$, $t$, $Q^2$, $E_{b}$, $\varphi$) as well as the value and experimental error of the cross section. In total, we are working with 3,862 unpolarized and 3,884 polarized data points.
Because these measurements are typically taken by sweeping across $\varphi$ values in a fixed kinematic bin, most points are unique only in $\varphi$, meaning there are relatively few values of $x_{Bj}, t, Q^2, E_{b}$. 
This presents an interesting challenge for ML approaches, which become much more capable of predicting across $\varphi$ values than they are of extrapolating these patterns to distant kinematic space. 

To help measure this effect, we use a $70\%/20\%/10\%$ split to separate the kinematic bins into three groups, which we'll call the training set, validation set and test set, respectively. Note that this differs from the standard ML practice of splitting the datapoints themselves into random groups. The training set is used to directly adjust the parameters of the model. 
The validation set is used as an indicator of performance during training - either for early stopping (see Sec \ref{sec:femtonet}), or for hyperparameter selection\footnote{A hyperparameter is a fixed constant that controls some aspect of an algorithm's behavior. They can have a significant impact on performance, but cannot be learned like many of the model's core parameters. As an example, SGD optimizes the parameters of a model, but the initial learning rate and batch size of the optimizer are predetermined hyperparameters.}. Final performance will be reported on the test set, which is made up of kinematic bins with which the model has never been trained or evaluated. A model that performs well on the test set is capable of predicting entire $\varphi$ curves in unseen regions of kinematic space, under the assumption that we are testing on inputs drawn from the same distribution as our training data. This assumption often breaks down in practice, as we'll see in Section \ref{sec:pseudo}. Generalizing to unseen kinematic bins is a more difficult challenge than generalizing to held-out \textit{datapoints} in bins that were part of the training set, because the overall shape of the $\varphi$ curve can be quite predictable as long as nearby points can provide a good frame of reference.

Based on the analysis of \citep{Kriesten:2019jep}, we experimented with transforming the dataset of ($x_{Bj}$, $t$, $Q^2$, $E_{b}$, $\varphi$) by concatenating (sin($\varphi$), cos($\varphi$), sin($2\varphi$), cos($2\varphi$), sin($3\varphi$), cos($3\varphi)$). These harmonics could serve as additional features that assist the model in learning the cross section's $\varphi$ dependence. We'll refer to the use of this additional input information as ``Harmonic Features."

We also scale the cross section values by 1000 to mitigate floating-point underflow and stabilize training. Unless otherwise noted, the plots below have been re-scaled to their original units.


\vspace{0.5cm}

\subsection{Baselines}
First, we examine the possibility of applying basic regression techniques to cross section fits to get an understanding of the difficulty of the problem and establish a performance baseline. We use the implementations provided in \cite{scikit-learn}:
\begin{itemize}
    \item \texttt{Linear} is a simple linear regression model. A vector of weights $\mathbf{W}$ and a scalar bias $b$ map the input kinematics, $x$, to the cross section prediction with $\hat{\sigma} = \mathbf{W}x + b$. These parameters are the solution to a closed-form system of equations that minimizes the mean squared error (MSE) of the model's predictions \cite{kenney1962linear}.
    \item \texttt{SVR} is a Support Vector Machine Regression model \cite{vapnik1999overview}. The model parameters, $\mathbf{W}$ and $b$ are the solution to the optimization problem \cite{chang2011libsvm}:
    \begin{align*}
        \mathop{min}_{\mathbf{w}, b, \gamma, \gamma^*} & \frac{1}{2}\mathbf{w}^T\mathbf{w} + C\sum_{i=1}^{l}\gamma_i + C\sum_{i=1}^{l}\gamma^*_i \\
        \textrm{subject to} \quad &\mathbf{w}^T\rho(\mathbf{x}_i) + b - \sigma_i \leq \epsilon + \gamma_i, \\
        & \sigma_i - \mathbf{w}^T\rho(\mathbf{x}_i) - b \leq \epsilon + \gamma^*, \\
        & \gamma_i, \gamma^*_i \geq 0, i = 1, \dots, l.
    \end{align*}
    Where $C$ and $\epsilon$ are hyperparameters that control error tolerance and $\rho$ is a kernel function that is used to map input vectors to a space where they may be more linearly separable. We experiment with both polynomial and radial basis function kernels.

\end{itemize}

\subsubsection{Evaluating Performance}
\label{sec:baselines}
Because of the large range of cross section targets present in our dataset, common regression metrics like mean squared error are biased towards a small number of kinematic bins. Therefore, we focus on statistics that normalize for the value of the true cross section. The mean absolute percentage error is a better comparison, but suffers from outliers when the model is inaccurate on very small cross sections. Instead, we compute the median absolute percentage error (MAPE). While this is an improvement, the data set has enough noise that relatively accurate fits can have high error. With this in mind, we also compute an accuracy metric, which we define to be the percentage of predictions that are close enough to the target value to be within the experiment's error bars defined according to Fig. \ref{fig:accuracy}.

\begin{figure}[!ht]
\begin{eqnarray*}
\text{Accuracy} = \frac{1}{N}\sum_{i}\Delta_{\sigma_{i}}
\end{eqnarray*}
\begin{eqnarray*}
\Delta_{\sigma_{i}} = \begin{cases}
0, \,\,\, \text{if}   &\begin{cases} \sigma_{i} - \hat{\sigma}_{i} > 0 \,\,\, \text{and}  \,\,\, |\sigma_{i} - \hat{\sigma}_{i}| > \delta_{\hat{\sigma}_{i}} + \delta_{\sigma_{i, \uparrow}}\\
\sigma_{i} - \hat{\sigma}_{i} < 0 \,\,\, \text{and}  \,\,\, |\sigma_{i} - \hat{\sigma}_{i}| > \delta_{\hat{\sigma}_{i}} + \delta_{\sigma_{i, \downarrow}}\\
\end{cases} \\
\\
1, \,\,\, \text{else}   &\begin{cases} \sigma_{i} - \hat{\sigma}_{i} > 0 \,\,\, \text{and}  \,\,\, |\sigma_{i} - \hat{\sigma}_{i}| < \delta_{\hat{\sigma}_{i}} + \delta_{\sigma_{i, \uparrow}}\\
\sigma_{i} - \hat{\sigma}_{i} < 0 \,\,\, \text{and}  \,\,\, |\sigma_{i} - \hat{\sigma}_{i}| < \delta_{\hat{\sigma}_{i}} + \delta_{\sigma_{i, \downarrow}}\\
\end{cases} \\
\end{cases}
\end{eqnarray*}
\caption{We use an accuracy metric which is defined by the number of points that lie within experimental error bars. $\Delta_{\sigma_{i}}$ is assigned a value of +1 or 0 based on the difference between the cross section prediction, $\hat{\sigma}_{i}$, and the data, $\sigma_{i}$. The assignment is determined by a flowchart of criteria. First we determine whether the difference is positive or negative, indicating whether we are looking at the positive side of the error bar or the negative. Then we compare the absolute value of the difference $|\sigma_{i} - \hat{\sigma}_{i}|$ to the size of the experimental error, $\delta_{\hat{\sigma}_{i}}$, plus asymmetric systematic errors, $\delta_{\sigma_{ i, \uparrow}},\delta_{\sigma_{i, \downarrow}}$. If the absolute difference is less than the size of the error bar then we assign a +1 value to $\Delta_{\sigma_{i}}$, else we assign a 0. The accuracy is then defined as the sum of all $\Delta_{\sigma_{i}}$ divided by total number of data points.}
\label{fig:accuracy}
\end{figure}

After a hyperparameter search, the best model is selected according to its MAPE on the validation set; it is then evaluated for the first time on the test set. The results are shown in Table \ref{tab:baselines_results}. \\

\begin{table*}
\centering
\scriptsize
\arrayrulecolor{black}
\begin{tabular}{|l|r!{\color[rgb]{0.8,0.8,0.8}\vrule}r|r!{\color[rgb]{0.8,0.8,0.8}\vrule}r|r!{\color[rgb]{0.8,0.8,0.8}\vrule}r|r!{\color[rgb]{0.8,0.8,0.8}\vrule}r|} 
\arrayrulecolor[rgb]{0.8,0.8,0.8}\cline{1-1}\arrayrulecolor{black}
\hline
\hline
\multicolumn{1}{|c|}{} & \multicolumn{4}{c|}{UU}                                                                                                                                                                                               & \multicolumn{4}{c|}{LU}                                                                                                                                                                                                \\ 
\arrayrulecolor[rgb]{0.8,0.8,0.8}\cline{1-1}\arrayrulecolor{black}\cline{2-9}
\multicolumn{1}{|c|}{} & \multicolumn{2}{c|}{Standard}                                                                             & \multicolumn{2}{c|}{Harmonic Features}                                                                    & \multicolumn{2}{c|}{Standard}                                                                             & \multicolumn{2}{c|}{Harmonic Features}                                                                     \\ 
\hline
\multicolumn{1}{|c|}{Method}                           & \multicolumn{1}{l!{\color[rgb]{0.8,0.8,0.8}\vrule}}{Median \% Error} & \multicolumn{1}{l|}{Accuracy (\%)} & \multicolumn{1}{l!{\color[rgb]{0.8,0.8,0.8}\vrule}}{Median \% Error} & \multicolumn{1}{l|}{Accuracy (\%)} & \multicolumn{1}{l!{\color[rgb]{0.8,0.8,0.8}\vrule}}{Median \% Error} & \multicolumn{1}{l|}{Accuracy (\%)} & \multicolumn{1}{l!{\color[rgb]{0.8,0.8,0.8}\vrule}}{Median \% Error} & \multicolumn{1}{l|}{Accuracy (\%)}  \\ 
\hline
\hline
Linear                                                 & 238.87                                                               & 2.10                                & 343.68                                                               & 1.0                                & 293.03                                                               & 19.68                              & 333.23                                                               & 22.34                               \\ 
\hline
SVR                                                    & 45.73                                                                & 19.14                              & 37.57                                                                & 27.13                              & 68.09                                                                & 57.71                              & 72.97                                                                & 57.98                               \\ 
\hline
Theory &13.39 & 49.50 & N/A & N/A & 62.46 & 62.67 & N/A & N/A \\
\hline
\textbf{FemtoNet}                                      & \textbf{9.99}                                                          & \textbf{61.97}                     & N/A                                                                  & N/A                                & \textbf{56.06}                                                       & \textbf{63.03}                     & N/A                                                                  & N/A                                 \\
\hline \hline
\end{tabular}
\caption{Results of baseline methods on the Test set. Each training run uses an identical Train/Val/Test split. We focus on median absolute percentage error because the large range of cross section sizes biases metrics like mean squared error towards a small number of kinematic bins. Accuracy is the \% of predictions that lie within experimental error bars. Best results in bold for clarity.}
\label{tab:baselines_results}
\end{table*}

Overall, the baselines perform poorly. While the linear models' high error is hardly a surprise given the cross section's trigonometric shape across $\varphi$, the SVR model with access to high degree polynomial features still struggles to find an accurate fit. Results on the use of additional harmonic features are mixed, and we decide against using them in the experiments that follow. In general, high error metrics like these highlight the complexity of the cross section function across the whole range of kinematic bins. Simple approaches are not sufficient for accurate predictions. 

\subsection{Training a Deep Network on DVCS Data}
\label{sec:femtonet}

\begin{figure}[!htb]
\label{loss-curves}
\minipage{0.5\textwidth}
  \includegraphics[width=\linewidth]{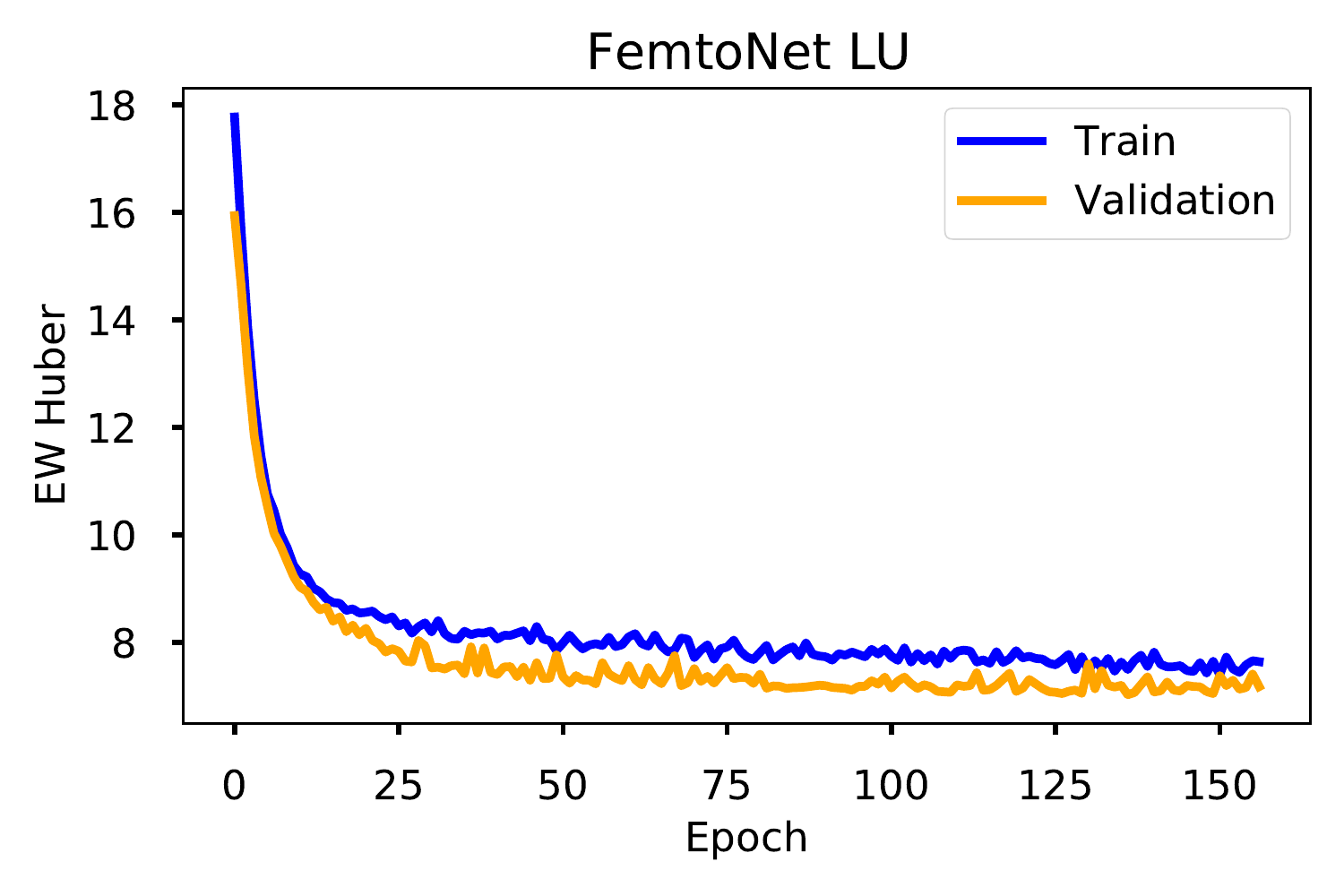}
\endminipage\hfill
\minipage{0.5\textwidth}
  \includegraphics[width=\linewidth]{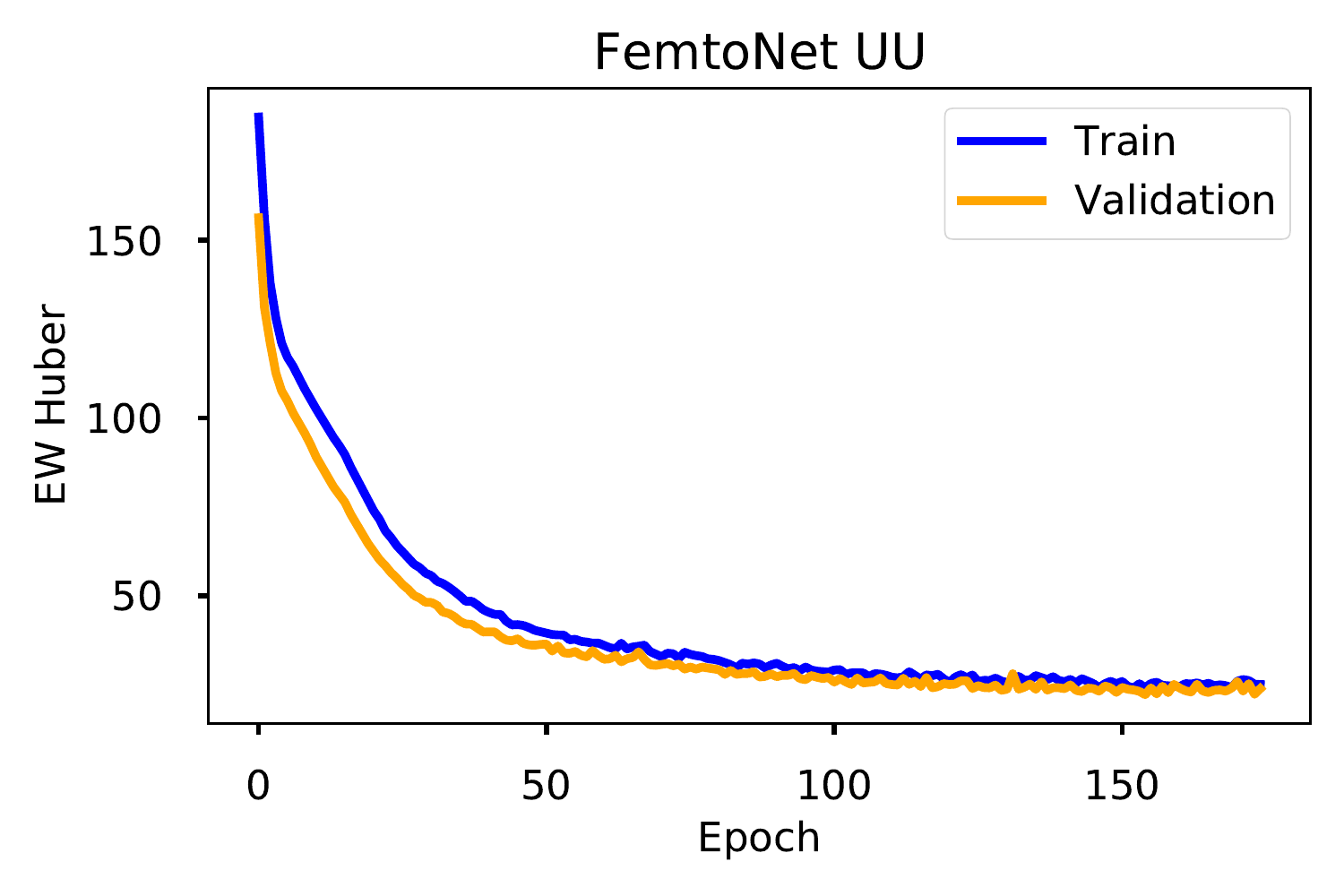}
\endminipage\hfill
\caption{Training curves for FemtoNet LU (top) and UU (bottom). The train loss is computed across batches of the training set while the validation loss is measured at the end of each epoch. When the validation loss has not hit a new minimum in the last 20 epochs, training ends to avoid overfitting.}
\end{figure}

We train two separate NNs, one to predict $\sigma_{UU}$ and the other to predict $\sigma_{LU}$. There are several major challenges in applying Deep Learning to this problem. The first is the range of the cross section data values; the target values for our networks' predictions span seven orders of magnitude in both the polarized and unpolarized cross sections. While the last layer of our architecture uses a linear activation function that can output any scalar, this wide range in targets destabilizes training. When kinematic points with large cross sections are sampled in a batch of training data, they can create outlier loss values that increase the size of our update gradients and make large adjustments to model parameters that decrease performance in regions with a smaller cross section. This effect is magnified by the MSE function, which is quadratic in the absolute error of our models' predictions. We address this by changing our loss function to the Huber loss:
\begin{align}
        \mathcal{L}_{\theta}^{H}(\textbf{x}, \sigma) =
        \begin{cases}
        \frac{1}{2}(\sigma - f_{\theta}(\textbf{x}))^{2},  & \text{if} \left|\sigma - f_{\theta}(\textbf{x})\right| \leq \delta ;\\
        \\
        \frac{1}{2}|\sigma - f_{\theta}(\textbf{x})|,  & \text{if} \left|\sigma - f_{\theta}(\textbf{x})\right| > \delta ; \\
        \end{cases}
\label{eq:huber}
\end{align}
The Huber loss is quadratic in the model's error when that error is less than a hyperparemter $\delta$, and linear otherwise. This limits the effect of large cross section values relative to MSE. The median value of the cross section data is relatively small, so the squared error term is stable for most datapoints and therefore the specific choice of $\delta$ is unimportant; we fix $\delta = 1$. 
 Many data points have error bars that are listed well above $100\%$ of the cross section value. Fitting to the center of such a wide range can create irregularities that generalize poorly and encourage overfitting. Ideally, our model would pay less attention to highly uncertain measurements. Towards this goal, we consider a variation on the standard Huber Loss that incorporates the relative size of the experimental error bars for each datapoint. Let $\Delta\sigma$ represent the size of the symmetric error bar, while $\Delta\sigma_{\downarrow}$ and $\Delta\sigma_{\uparrow}$ represent the negative and positive asymmetric error, respectively. The Error Weighted Huber Loss (EW Huber) is then defined:
 \begin{align}
      \mathcal{L}_{\theta}^{EW}(\textbf{x}, \sigma) =  \frac{\mathcal{L}_{\theta}^{H}(\textbf{x}, \sigma)}{1 + ((2 \Delta\sigma + \Delta\sigma_{\downarrow} + \Delta\sigma_{\uparrow}) / \sigma)}
      \label{eq:ewhuber}
 \end{align}
 The EW Huber discounts the loss value of datapoints where the total size of the error bar is large relative to the scale of the cross section. Points with no experimental error stay unchanged and the effect of the weighting increases with the experimental uncertainty.
 
\begin{figure}
    \centering
    \includegraphics[scale=.35]{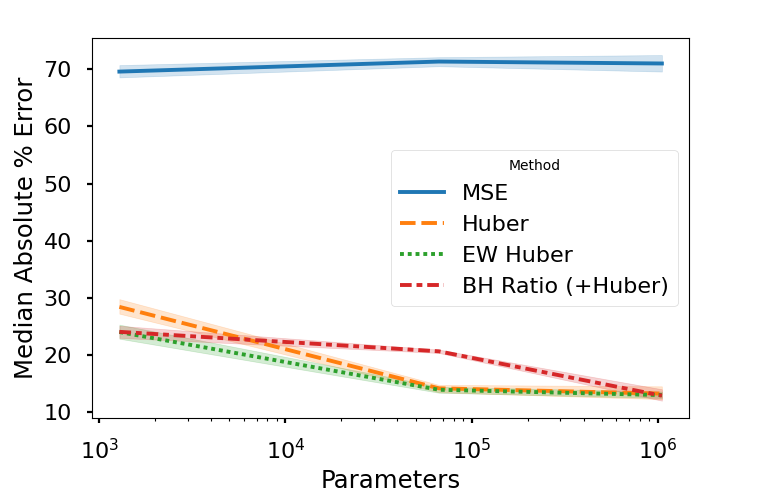}
    \caption{Comparing several of our design choices across varying model sizes. 'MSE' uses a standard Mean Squared Error loss functions. `Huber` switches to a clipped variant of MSE that protects against outliers. `BH Ratio` generates cross sections by multiplying the output of the network by the value of the Bethe Heitler component, which is calculated based on known theory \citep{Kriesten:2019jep} inside of a custom layer. `EW Huber` reweights terms in the huber loss based on the size of the experimental error bars - reducing the importance of imprecise datapoints. Gradient clipping is used across all experiments.}
    \label{fig:tricks}
\end{figure}

 In many regions of kinematic space, the unpolarized cross section is dominated by the Bethe-Heitler component, which can be computed as a known function of the network's inputs. We take advantage of this by implementing the Bethe-Heitler calculation as a custom neural network layer with fixed parameters that cannot be adjusted by gradient descent. This layer computes the Bethe-Heitler component of the input point, which is then multiplied by the output of the standard NN architecture to produce the predicted cross section. This effectively changes the function the NN must learn to approximate from $\sigma$ to the simpler $\sigma /BH$.
 We test these approaches on increasingly large NN sizes to see how they affect performance, and the results are shown in Figure \ref{fig:tricks}. The Huber loss turns out to be essential for successful training. Performance improves with model size - perhaps the biggest advantage of Deep Learning techniques. The BH Ratio simplifies the NN function enough to help small models, but as models grow larger their representational capacity improves and this simplification becomes unnecessary.

\subsubsection{Overfitting and Early Stopping} Our models' goal is to make accurate predictions in previously unseen kinematic bins. By improving their performance on the training set through gradient descent, we tend to improve their performance on the test set. However, over-optimizing on those training points can eventually come at the expense of test set performance. This effect is known as `overfitting' and is an important concern when training deep networks. A simple and effective heuristic to stop training before overfitting occurs is to periodically evaluate the model on the \textit{validation} set. When the validation loss stops decreasing, we are in danger of overfitting, and end the training run. This approach, known as Early Stopping, uses the validation set as a stand-in for the test set. SGD assumes the training, testing, and validation sets are sampled from the same underlying data distribution. Therefore, we can use the performance on the validation set as a sample of our performance on the testing set, as long as we aren't making gradient updates on the validation data. We use Early Stopping in every training run, ending training after the validation loss has not decreased in more than 20 full iterations through the training set (epochs).

\begin{figure}
\includegraphics[width = 8.6 cm, height = 4.7 cm]{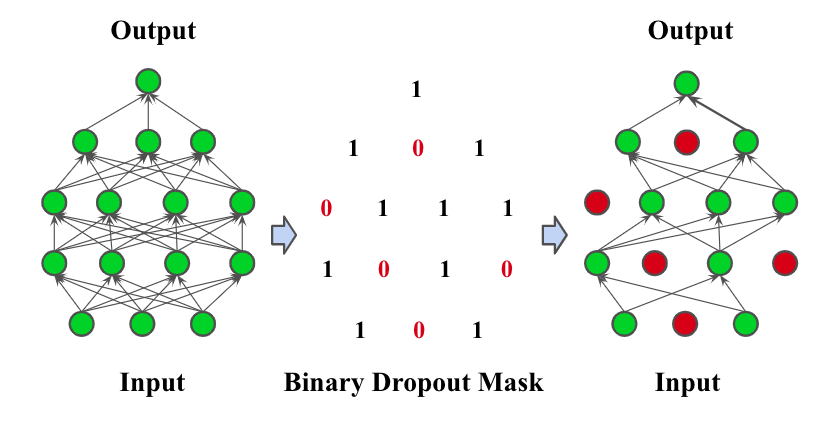}
\caption{Example of the application of a binary dropout mask to a fully connected neural network. The 1's and 0's are assigned based on a probability distribution over each weight. When we re-sample the binary mask over each forward pass of the network, the variance of the output can be used as an approximation of statistical uncertainty of the neural network predictions. }
\label{fig:drop_pic}
\end{figure}

 The size of the dataset is another obstacle. While many of the most successful applications of DL use training sets of millions \citep{imagenet_cvpr09} or even billions \citep{brown2020language} of samples, the DVCS experiments at our disposal leave us with a little under $4,000$ datapoints. Small datasets increase the risk of overfitting and often force us to \textit{regularize} the model, or to restrict it's ability to move freely in parameter space and arrive at a solution that is hyper-specific to the training set. We accomplish this with Dropout regularization \citep{srivastava2014dropout}. With Dropout, connections between neurons in the network are severed before each forward pass with some probability $p$, which becomes an important hyperparameter. This is typically implemented by sampling binary weight matrices that serve as masks of the network weights - setting them to zero when the connection should be removed and one otherwise. This is shown pictorially in Fig. \ref{fig:drop_pic}. Intuitively, Dropout reduces overfitting by preventing the network from being overly reliant on the outputs of a small number of neurons. There are many other approaches for regularization. We focus on Dropout because of it's use in predicting model uncertainty, which will be discussed further in Sec \ref{sec:4}.
 
\subsubsection{FemtoNet} The final architectures of our cross section prediction networks are determined with by a large search, guided by the Hyperband algorithm \citep{li2018hyperband}. Hyperband trains many possible variations of the network architecture and uses the results to decide which kinds of networks it should dedicate more time/resources to investigating and which kinds to disregard. The final UU and LU models both use the Adam optimizer \citep{kingma2014adam} with learning rate $7.5\mathrm{x}10^{-4}$. The dropout probability $p = .25$. FemtoNet UU uses two hidden layers, both of size $2048$, while FemotoNet LU uses two hidden layers with sizes of $2048$ and $1024$. Both networks use the EW Huber Loss. They also normalize inputs before the first layer, with mean and std statistics determined from the training set.

\begin{figure}[!ht]
\label{fig:nnarch}
\includegraphics[scale=0.4]{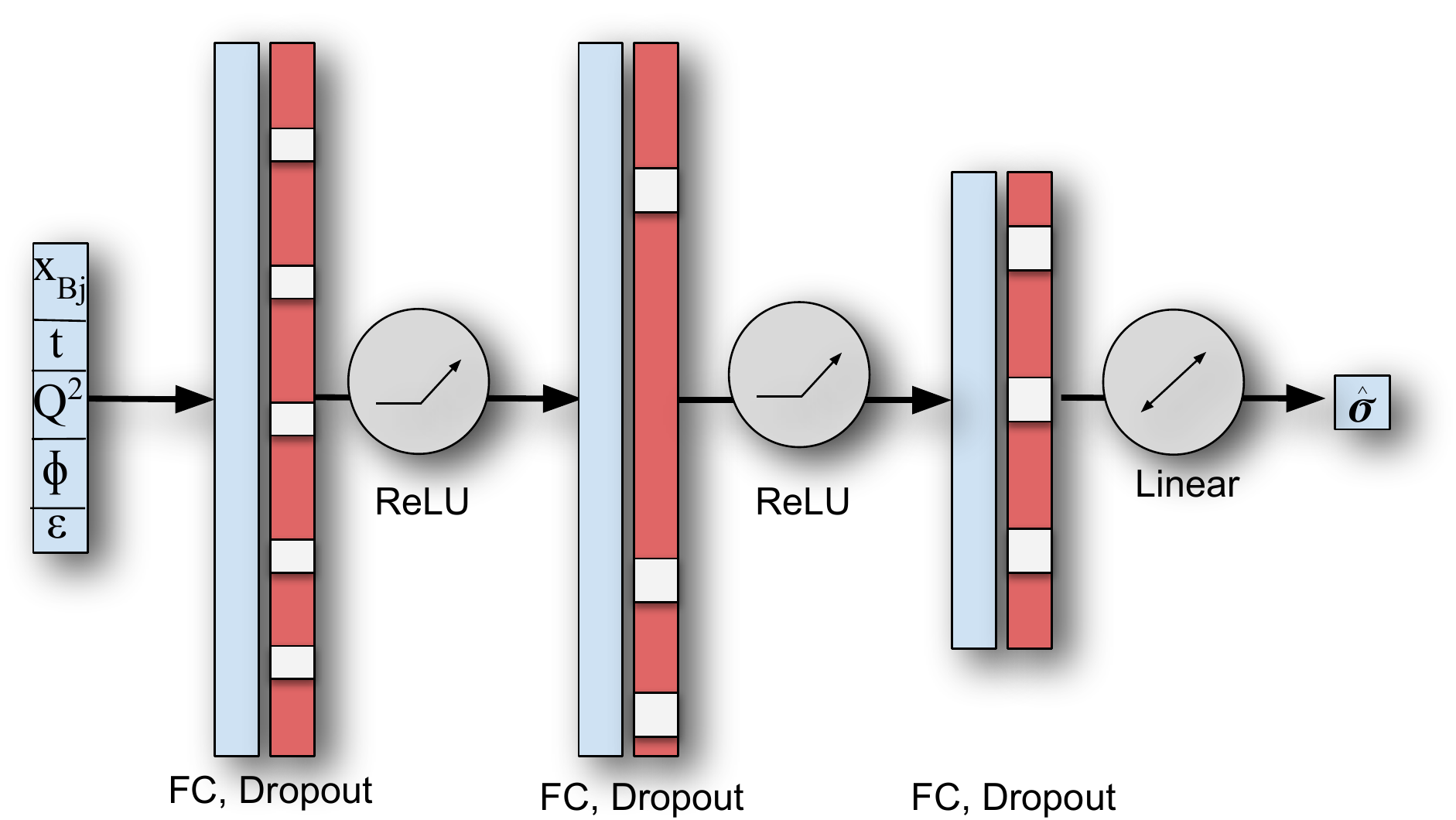}
\caption{A schematic of the FemtoNet model architecture. Two large hidden layers with ReLU activations feed into a linear output layer that determines the cross section prediction. Dropout is applied after each Fully Connected (FC) layer.}
\end{figure}

We evaluate FemtoNet on the same test set as the baseline approaches, with the results also listed in Table \ref{tab:baselines_results}. FemtoNet comfortably outperforms the ML baselines, though the improvement is much more significant in the unpolarized cross section than the polarized data. We discuss this finding further in Sec \ref{sec:uu_vs_lu}. In addition to comparing against other ML techniques, we compare against the theoretical cross section model developed in \citep{Kriesten:2019jep}. The NN is even more accurate over the test set - adding DVCS cross section prediction to a long list of problems in ML in which the performance of hand crafted or manually specified heuristics or formulae can be matched by \textit{learned} features that are automatically discovered. However, the theory model has the important advantage of consistency across kinematics far outside the region of the training data, while the NN is less constrained.

\begin{center}
\begin{figure}[!ht]
\begin{tabular}{cc}
  \includegraphics[scale=0.4]{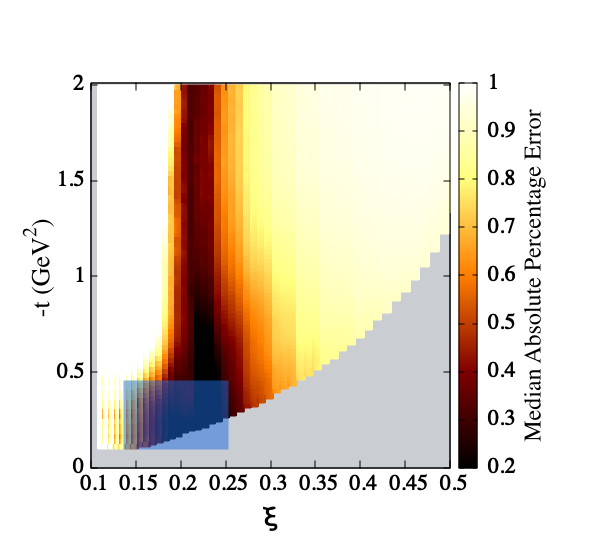} \\
  \includegraphics[scale=0.4]{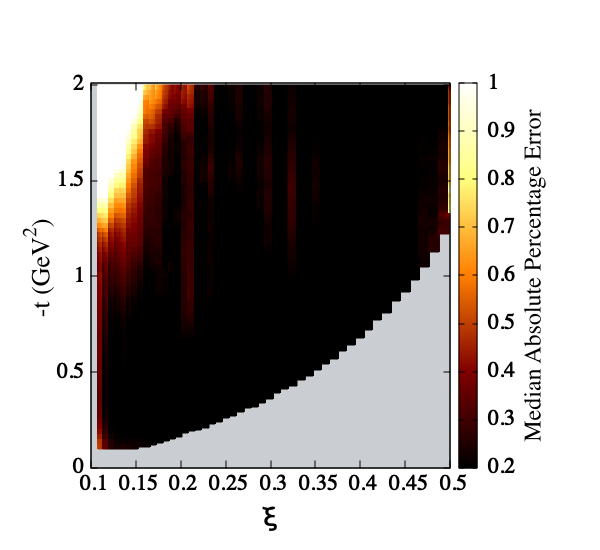} \\
\end{tabular}
\caption{Agreement between the theory-based cross section model and the NN approximation before ({\it{Top}}) and after ({\it{Bottom}}) the addition of generated pseudo-data. This is calculated fixing $Q^{2} = 2$ GeV$^{2}$ plotted as a function of the variables momentum transfer variables $\xi$ and $t$. The shaded rectangle represents where current DVCS exclusive data sit at a $Q^{2} \approx 2$ GeV$^{2}$.} 
\label{fig:heatmap}
\end{figure}
\end{center}

In the upper plot of Figure \ref{fig:heatmap}, we compare the cross section predictions of FemtoNet to the theoretical predictions across a slice of $t$ and $\xi$. The region of existing experimental data is indicated by the shaded rectangle. Unsurprisingly, the two agree quite strongly in this area, where both have already been proven to be accurate based on the test set evaluation. However, as we move away from existing data we see a dramatic spike in the difference in the NN's predictions relative to the theory-based model. Note that this experiment says nothing about the accuracy of the theory's predictions over these kinematics - there is no experimental data here to serve as the ground-truth. All we can conclude is that the theory and ML significantly disagree. The ability of NNs to generalize outside their training distributions is an active area of research in ML. Deep Learning is often concerned with high dimensional data, such as images or audio files, where any concept of distance between datapoints is essentially meaningless. In contrast, the $x_{Bj}$, $t$, and $Q^2$ variables of kinematic space make it easy to think about ``how far" our model can generalize, and may make this problem of interest to the broader ML community. Fig \ref{fig:heatmap} suggests that FemtoNet learns a generalizable representation of the $t$ dependence, but struggles with other kinematic variables. 

\subsection{Scaling to Larger Datasets}
\label{sec:pseudo}
    While the cross section measurments used in our analysis consist of around 4,000 datapoints, that total will likely see significant growth from future experiments. We are interested in developing a method that can scale as new measurements are collected. This can be tested by expanding the kinematic region to include a larger range of $x_{Bj}, t$, and $Q^{2}$. In particular, we generate pseudo-data in regions outside the range where data currently exists to measure the network's compatibility with theoretical calculations of the cross section.

  \subsubsection{Pseudo data}

 \begin{center}
 \begin{figure}[!ht]
 \includegraphics[scale=0.296]{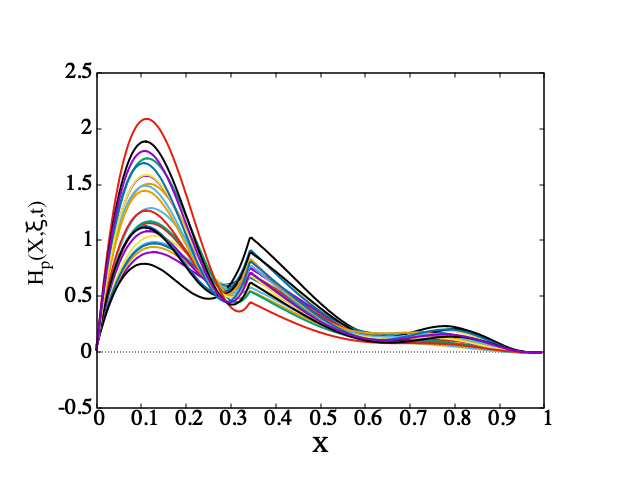}
 \includegraphics[scale=0.296]{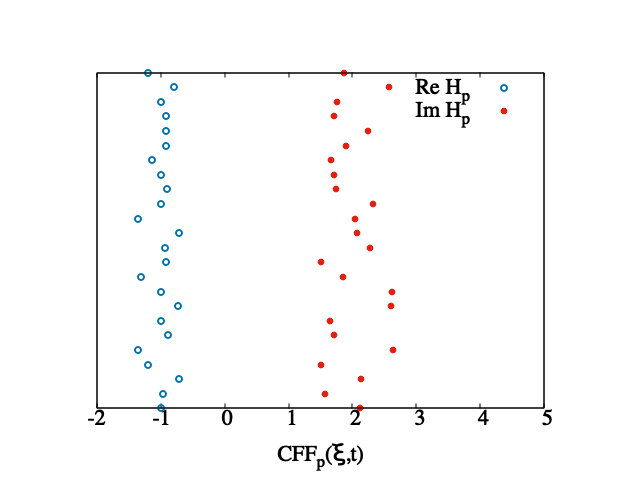} \\
 \includegraphics[scale=0.376]{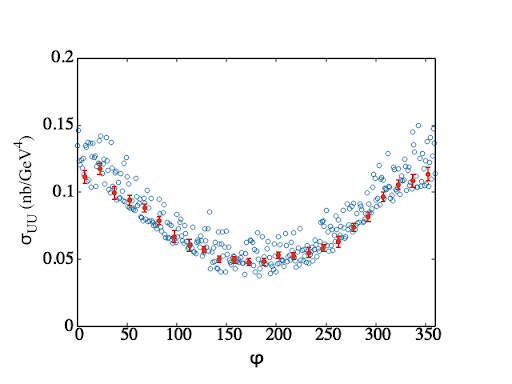}
 \includegraphics[scale=0.296]{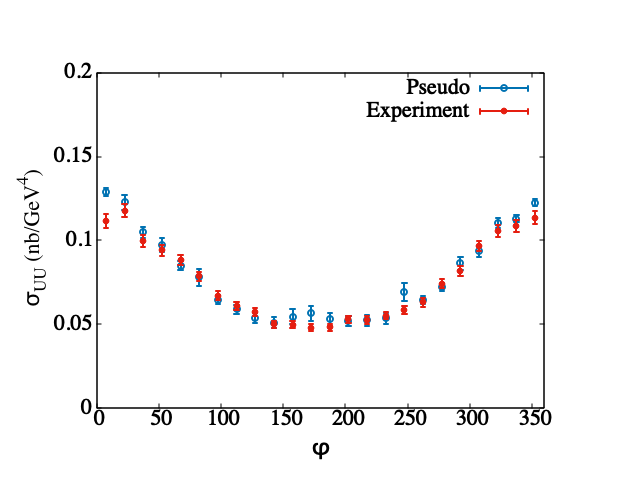}
 \caption{Process for creating pseudo-data at the kinematics $Q^{2} = 1.820 \,\, \text{GeV}^{2}, \,\, x_{Bj} = 0.343, \,\, t = -0.172 \,\, \text{GeV}^{2}, \,\, \text{E}_{b} = 5.75 \,\, \text{GeV}$. ( {\it{1$^{\text{st}}$}} a) construction of GPD envelopes, ({\it{2$^{\text{nd}}$}} b) generation of CFFs from GPD envelopes, ({\it{3$^{\text{rd}}$}} c) creation of an envelope of cross section values, ({\it{4$^{\text{th}}$}} d) final pseudo data with simulated errors.}
 \label{fig:pseudo_gen}
 \end{figure}
 \end{center}

In Figure \ref{fig:pseudo_gen} we demonstrate the process for generating pseudo-data using several diquark model calculations to parametrize the GPDs. The panels are described in descending order in detail in the text from subsections (a - d).

a) GPD envelopes are generated using two parametric forms following \cite{Goldstein:2012az} as $H_{VA}$, $E_{VA}$, $\widetilde{H}_{VA}$, $\widetilde{E}_{VA}$ and \cite{VGG:1999} as $H_{VGG}$, $E_{VGG}$, $\widetilde{H}_{VGG}$, $\widetilde{E}_{VGG}$. To ensure an unbiased and suitably ``random" choice of the parametric form for the GPD we vary the parameters $P_{i}^{o}$ of each model by a random fluctuation, $\Delta_{i}$, of up to $\pm$ 10 \% of $P_{i}^{o}$ dictated by a random number generator. We create the parameters of the model in the following way.
 
 \begin{equation}
 P_{i} = P_{i}^{o} + \Delta_{i}
 \end{equation}
 
\begin{figure}[h!]
 \centering
 \includegraphics[scale=0.36]{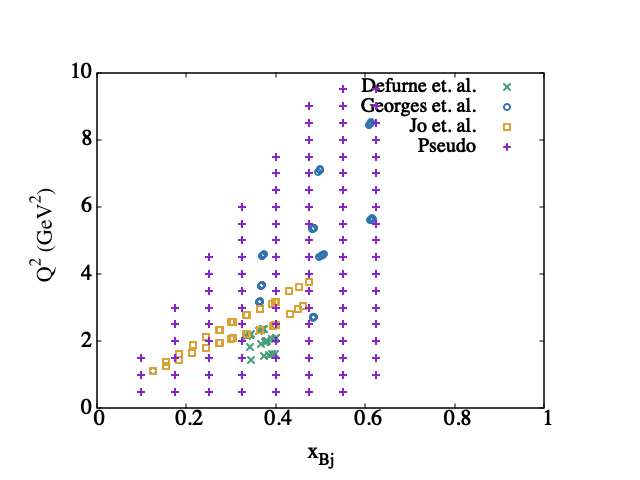}
 \includegraphics[scale=0.45]{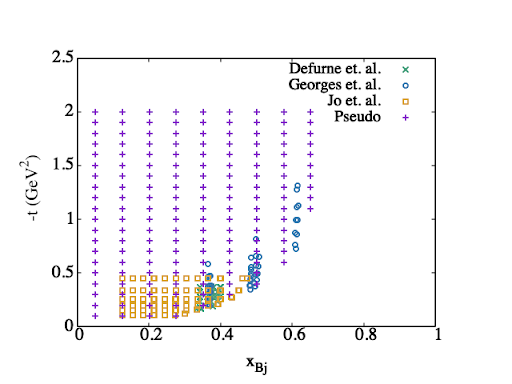}
 \caption{Kinematic regions in ({\it{Top}}) $x_{Bj}$ versus $Q^{2}$ and ({\it{Bottom}}) $x_{Bj}$ versus $t$ where we generate pseudo-data compared to where current DVCS data exists.}
 \label{fig:pd}
 \end{figure}

 This is done for both parametric forms of the GPDs. The final envelope is generated by combining the two models in a random combination,
 
 \begin{eqnarray}
 F &=& \alpha  \, F_{VA} + \beta  \, F_{VGG}
 \end{eqnarray}
 where $F = [H,E,\widetilde{H},\widetilde{E}]$,   $\alpha$ is a randomly generated number between 0 and 1, and $\beta = 1 - \alpha$. Using this method, for each kinematic bin we generate 25 GPD curves as shown in the top panel of Fig.\ref{fig:pseudo_gen}.
 
 b) We evolve the GPDs using pQCD evolution equations and convolute the GPDs with the corresponding Wilson coefficient function to calculate an envelope of 25 real and imaginary parts of the Compton form factors. The imaginary part of the Compton form factor is directly related to the value of the GPD at the cross over point between ERBL and DGLAP regions multiplied by $\pi$, therefore it is more susceptible to the random fluctuations of the GPDs. This is in contrast to the real part of the Compton form factor which is constrained by the enveloping Wilson coefficient function and integral over the longitudinal momentum variable $x$. 
 
 c) The CFF envelope enters into the cross section following the theory discussed in section \ref{sec:2} above and \cite{Kriesten:2019jep, Kriesten:2020wcx}. We generate pseudo-data by evaluating the cross section at azimuthal angle, $\varphi$, values starting at 7.5 and going in increments of 15 degrees until we reach 360. For each of the 24 $\varphi$ values we generate 25 cross section points using each of the 25 generated CFFs. This creates an envelope of cross section calculations for each kinematic bin. The spread of the cross section values can be seen to be suitable enough to be random, but it is approximately centered around the actual data values.
 
 d) For each $\varphi$ value of the data, we randomly pick one data point from the 25 in the cross section envelope. The random choosing of the data also simulates some of the statistical fluctuations naturally present in the experimental data. We generate the pseudo-data errors by comparing to the data set in the regions where there is data and controlling the error to match the relative size of the cross section errors. Using pseudo data, we have control over the size of the error bars giving us a unique opportunity to study the effects of experimental error when using a neural network analysis. We then use this metric to extend into regions where there is no data, simulating approximate errors in regions unexplored by experiment. The coverage of our pseudo-dataset relative to existing experimental data is shown in Figure \ref{fig:pd}.
 
 \subsubsection{Evaluating Performance on the Pseudo-dataset}
 
 We evaluate our Deep Learning approach on an expanded pseudo-dataset of more than $30,000$ unpolarized cross section measurements. One of the most important advantages of NNs is their scalability; it's not uncommon for hard problems in ML to be solved by simply scaling up a model's parameter count. This is certainly an option when fitting cross sections. However, for the sake of comparison, we use the same training process and model architecture developed for the experimental datasets. The results are shown in Table \ref{tbl:pseudo}. Our NN pipeline adapts well, highlighting its ability to learn an accurate approximation of the cross section across a large region of kinematic space where it's more difficult to rely on overfitting or `memorization.'

In the bottom of Fig. \ref{fig:heatmap}, we investigate the pseudo-data's impact on the shape of the NN's approximation relative to the theory-based model described in \citep{Kriesten:2019jep}. As discussed in Sec \ref{sec:femtonet}, the NN can diverge significantly from the theoretical model when predicting on kinematic bins that are far from existing measurements. When given access to cross sections in more distant regions of kinematic space (as part of the pseudo-dataset), this problem vanishes, underscoring the importance of collecting DVCS measurements across a wide range of kinematics in future experiments.

\begin{table}[]
\begin{tabular}{|c|c|c|}
\toprule
\hline
\hline
         & Median \% Error & Accuracy (\%) \\ \midrule \hline \hline
Linear   & 3090.56         & 1.50           \\ \midrule \hline
SVR      & 42.46           & 47.75         \\ \midrule \hline
\textbf{FemtoNet} & \textbf{12.50}            & \textbf{73.90}          \\ \bottomrule \hline \hline
\end{tabular}
\caption{Pseudo-dataset results. NN approaches scale well to a dataset of more than $30,000$ samples.}
\label{tbl:pseudo}
\end{table}

\section{Predicting the Cross Section}
\label{sec:4}

\subsection{Estimating Model Uncertainty During Inference}
\label{uncertainty-explanation}
When making predictions that have an effect on decisions or later calculations, we'd like to know the confidence level of our model - how sure are we that our predictions are accurate? This confidence is quantified as model uncertainty. Recent literature (see Refs.\cite{kumericki2011studying,moutarde2019unbiased} for DVCS, \cite{Ball:2017nwa,Ball:2011uy} and references therein for inclusive scattering) has adopted the ``replica method" as a means of estimating uncertainty. The replica method is similar to the Bootstrap aggregating \citep{breiman1996bagging} approach that is common in ML, in which a set of models is trained by sampling from the dataset with replacement and using the disagreement of their predictions to estimate uncertainty. The replica method consists of running the learning algorithm many times on versions of the dataset where points are randomly perturbed relative to their experimental error bars. This is meant to propagate the experimental error into the results of the model. After many cycles of resampling and retraining, we have a set of models trained on slightly different datasets. The mean output of this set of models is used as the prediction, and the standard deviation serves as an approximation of uncertainty. While this is a valid way to incorporate experimental/epistemic uncertainty into the modeling process, the repeated training can be prohibitively expensive when working with large models or datasets. Using multiple copies of the model parameters can be memory-intensive, difficult to vectorize, and less computationally efficient when making predictions. Ideally, we could recover model uncertainty from a single neural network.


One possibility is a Bayesian Neural Network, where we optimize a distribution over weights and sample from that distribution between each prediction \cite{pmlr-v37-blundell15}. While effective, these networks have significantly more parameters and can be more difficult to train. A simple alternative is to leave dropout turned on during inference. By resampling our binary mask over the weights between each forward pass, we create a prediction distribution that can be used to approximate model uncertainty. This technique can be shown \cite{gal2015dropout} to be an approximation of a Gaussian Process. Our model generates a batch of $k$ predictions over the same datapoint, each with different dropout masks. Unless otherwise noted, we set $k = 1000$. This smooths the uncertainty bands for plotting, but this value can be decreased if inference speed is a priority.

\subsection{Learning the $\varphi$ dependence}

\begin{center}
\begin{figure*}[!ht]
\begin{tabular}{cc}
  \includegraphics[width=72mm]{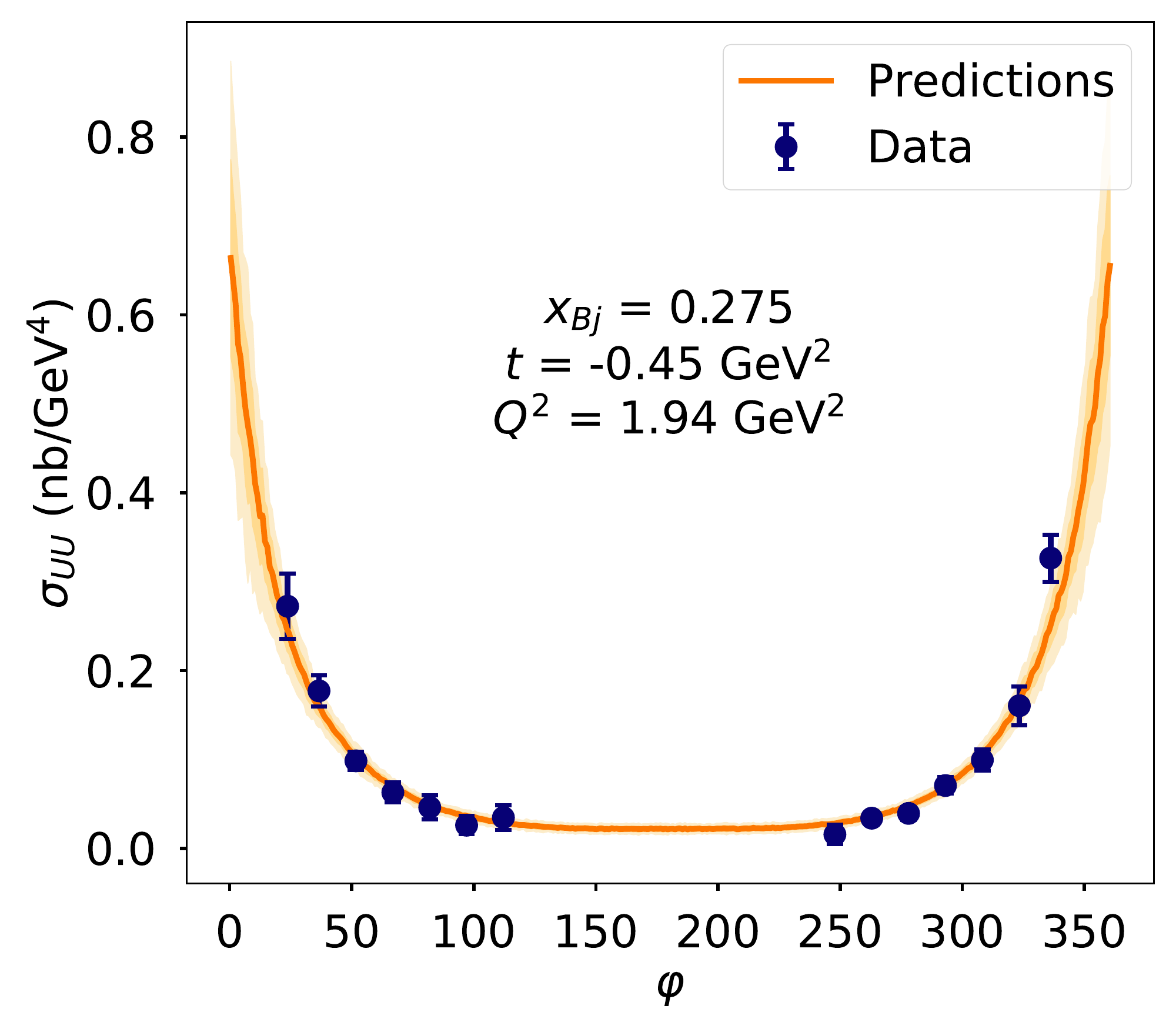} &   \includegraphics[width=72mm]{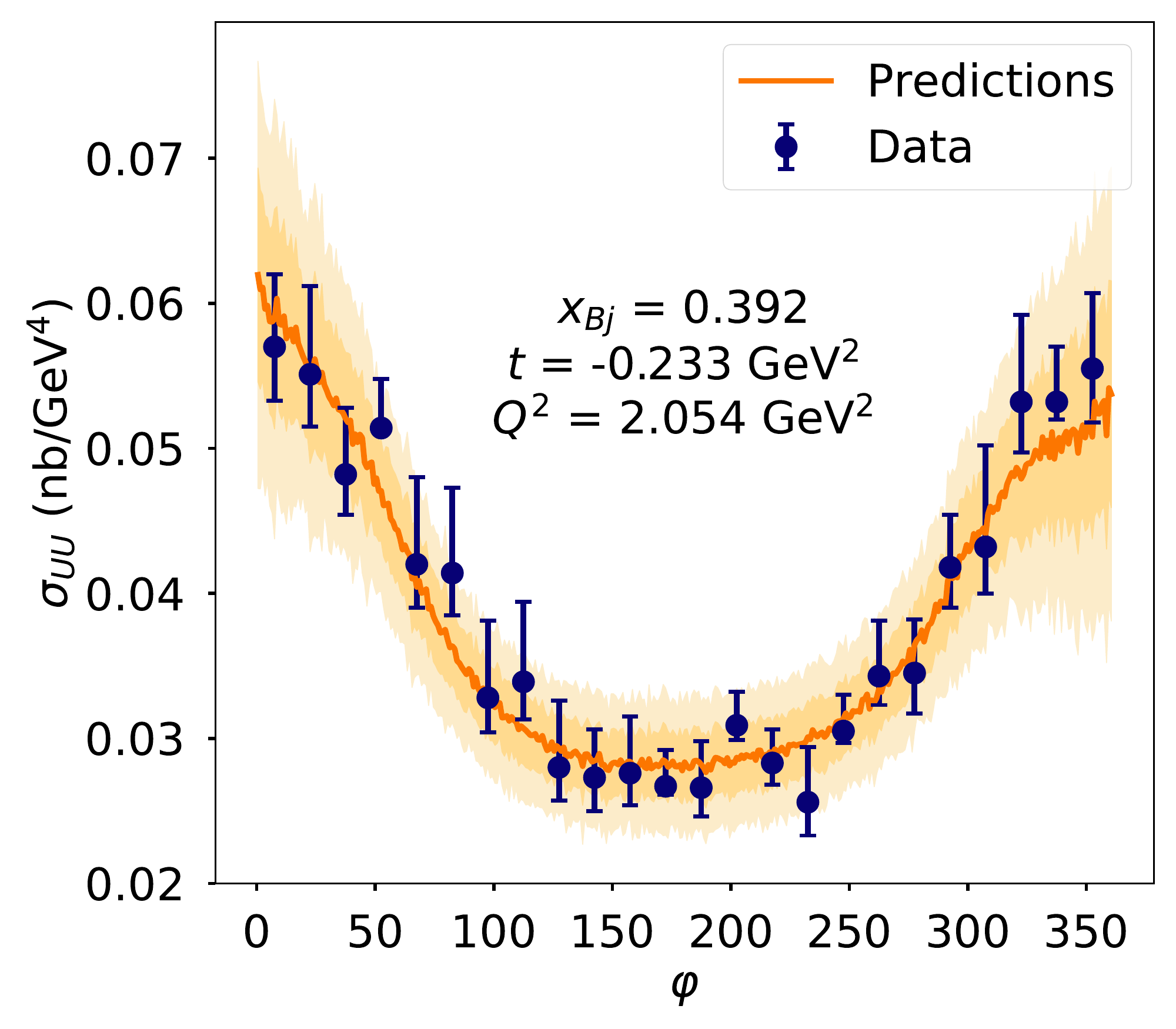} \\
 \includegraphics[width=72mm]{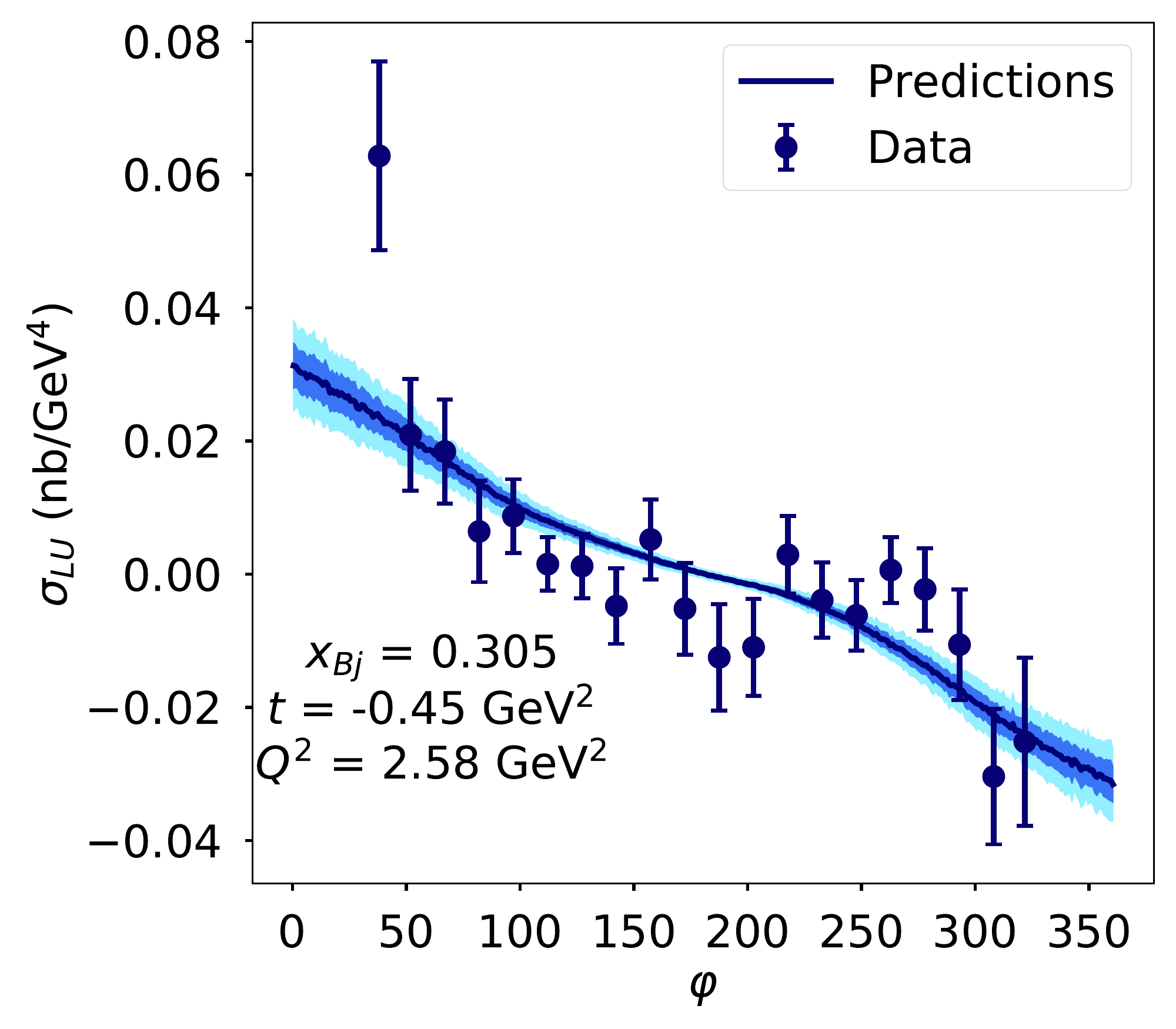} &   \includegraphics[width=72mm]{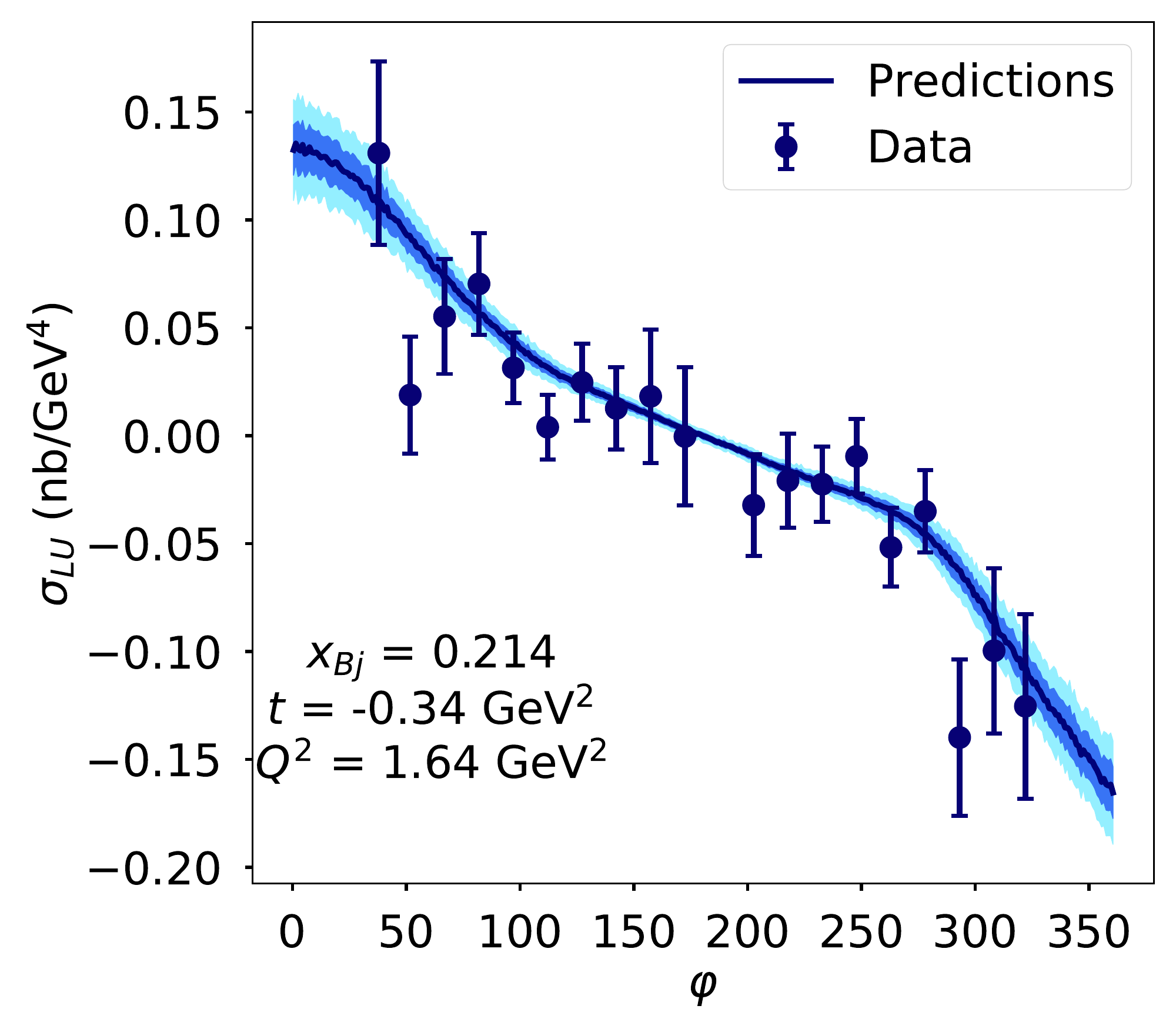} \\
\end{tabular}
\caption{The $\varphi$ dependence of $\sigma_{LU}$ (top row), and $\sigma_{UU}$ (bottom row) in previously unseen kinematic bins. Shading indicates two standard deviations of 1000 predictions.}
\label{fig:phi}
\end{figure*}
\end{center}

First, we examine the model's ability to fit the range of $\varphi$ values in existing kinematic bins ($x_{Bj}$, $t$, $Q^2$, $E_{b}$). FemtoNet's predictions in unseen kinematic bins from the test set are shown in Figure \ref{fig:phi}. The models learn to predict the cross section within experimental error bars, and are robust to noise and outliers, particularly in the Hall B datasets.

\subsection{Unpolarized vs. Polarized Results}
\label{sec:uu_vs_lu}

The total unpolarized cross section at leading twist contains 3 pieces as shown in Eq.(\ref{eq:xsx}): a piece that is bilinear in Compton form factors (DVCS), an interference piece that is linear in Compton form factors ($\mathcal{I}$), and a piece that has no dependence on Compton form factors and is an exactly calculated background parametrized by the elastic form factors (BH). The Bethe-Heitler cross section term, in most kinematic regimes, dominates the total cross section; often an order of magnitude or larger than the interference and DVCS terms combined. We can calculate the background to a well known degree of accuracy using covariant four vector products. The error and the noise of the cross section is thus due entirely to the terms which are parametrized by the Compton form factors. Since the background dominates the cross section, it then also dominates the noise inherent in scattering cross section data, effectively reducing the errors and simplifying the functional form. It is demonstrated in Table \ref{tab:baselines_results} that the neural network can predict the unpolarized cross section with a relatively small median percentage discrepancy.

This is in contrast to the polarized cross section which has a drastically different result. The functional form of the polarized cross section we find is more difficult for the neural network to learn. We can explain this by the presence of a large contribution of noise in the polarized data. By parity constraints, the longitudinally polarized beam on an unpolarized target (LU) cross section has no Bethe-Heitler background nor a term bilinear in Compton form factors at leading twist. It is linear in Compton form factors; as such, there is no error / noise suppression as in the unpolarized case. 

\begin{figure}[!ht]
\includegraphics[scale =0.35]{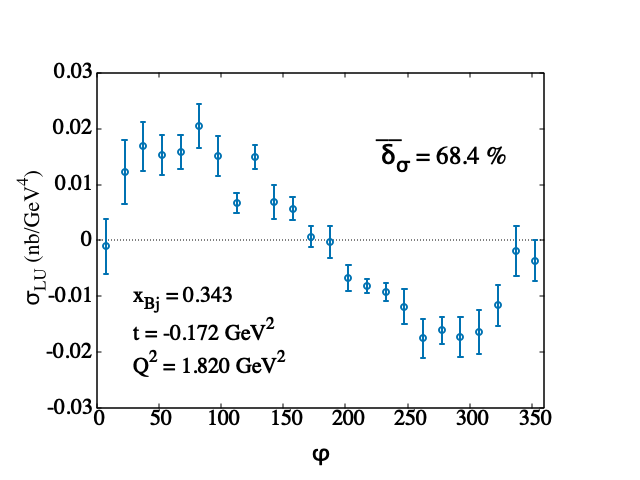} \\
\includegraphics[scale =0.35]{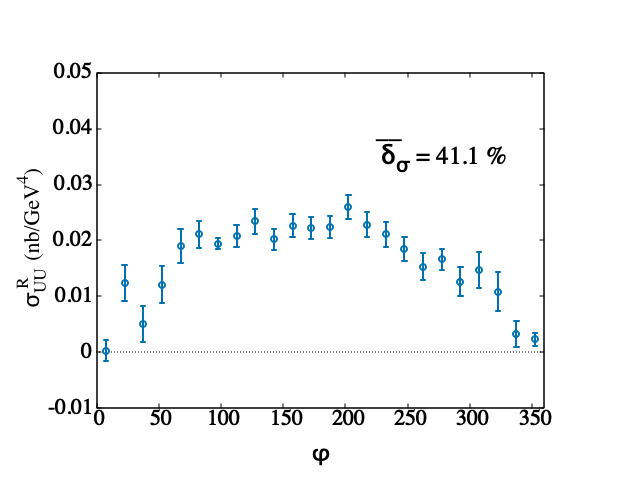}
\caption{({\it{Top}}) The total polarized cross section which at leading twist is linearly dependent on Compton form factors. ({\it{Bottom}}) The reduced unpolarized cross section which we define as the total cross section subtracting the Bethe-Heitler cross section. The reduced cross section has 2 terms both dependent on Compton form factors. Notice that the relative errors of these two converge without the presence of a background process in the unpolarized case. }
\label{fig:uuvlu}
\end{figure}

We can compare the LU polarized cross section to an equivalent reduced unpolarized cross section where we have subtracted off the Bethe-Heitler background to investigate the impact of the dominant term on the errors as shown in Figure \ref{fig:uuvlu}. Notice that the unpolarized cross section tends to converge to a relative error similar to that of the polarized cross section at similar kinematics. Where the relative error is defined as

\begin{eqnarray}
\overline{\delta}_{\sigma} = \frac{1}{N}\sum_{i} \frac{\delta_{\sigma_{i}}}{\sigma_{i}}
\end{eqnarray}

The background in the unpolarized cross section tends to simplify the functional form of the cross section while simultaneously smoothing out the noise and decreasing relative error. This, however, presents a problem for extracting Compton form factors where the cross section terms they parametrize tend to have large noise components. 

\subsection{Predictions in New Kinematic Regions}
\begin{figure}[!ht]
  \includegraphics[width=75mm]{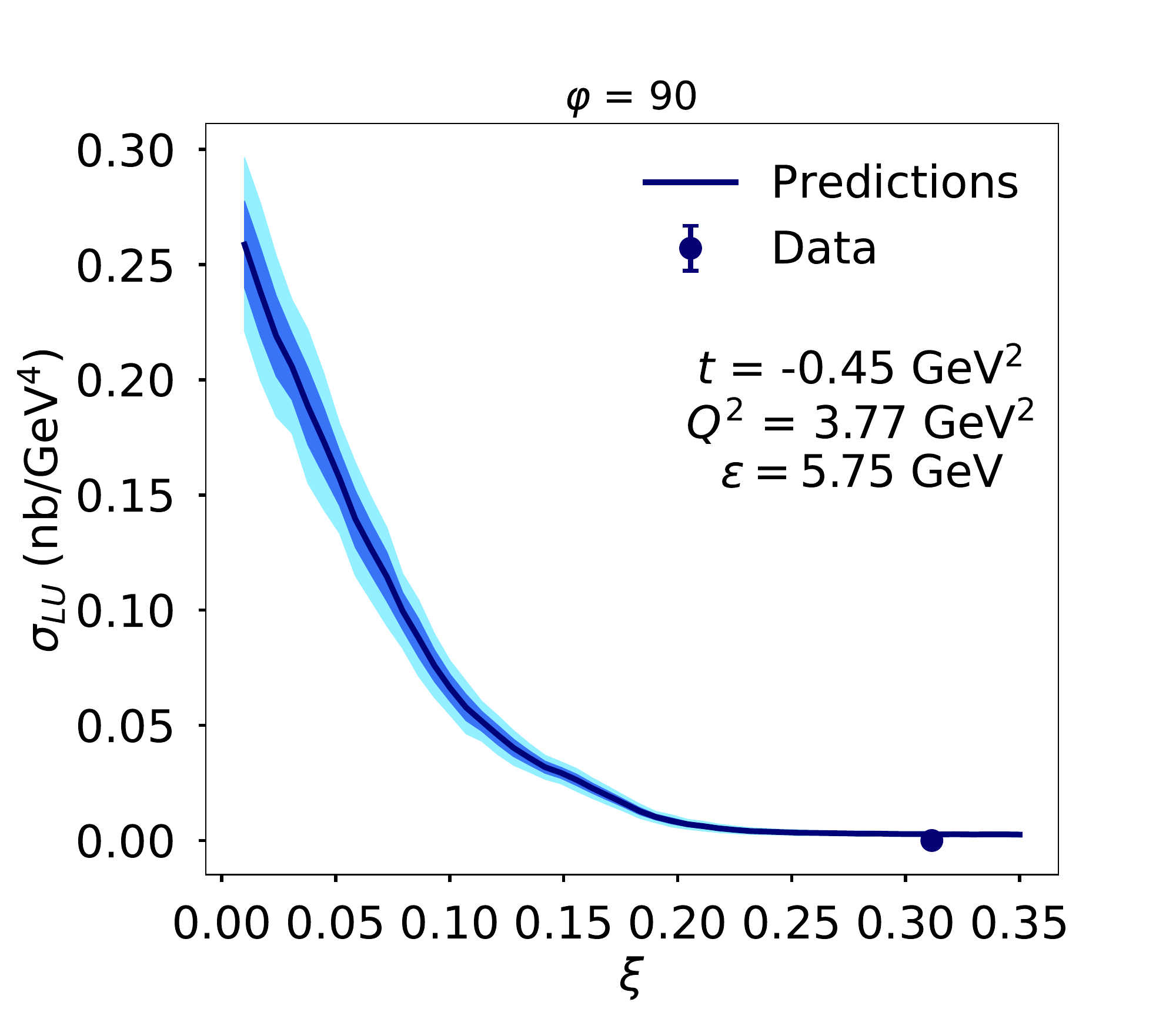}  \includegraphics[width=75mm]{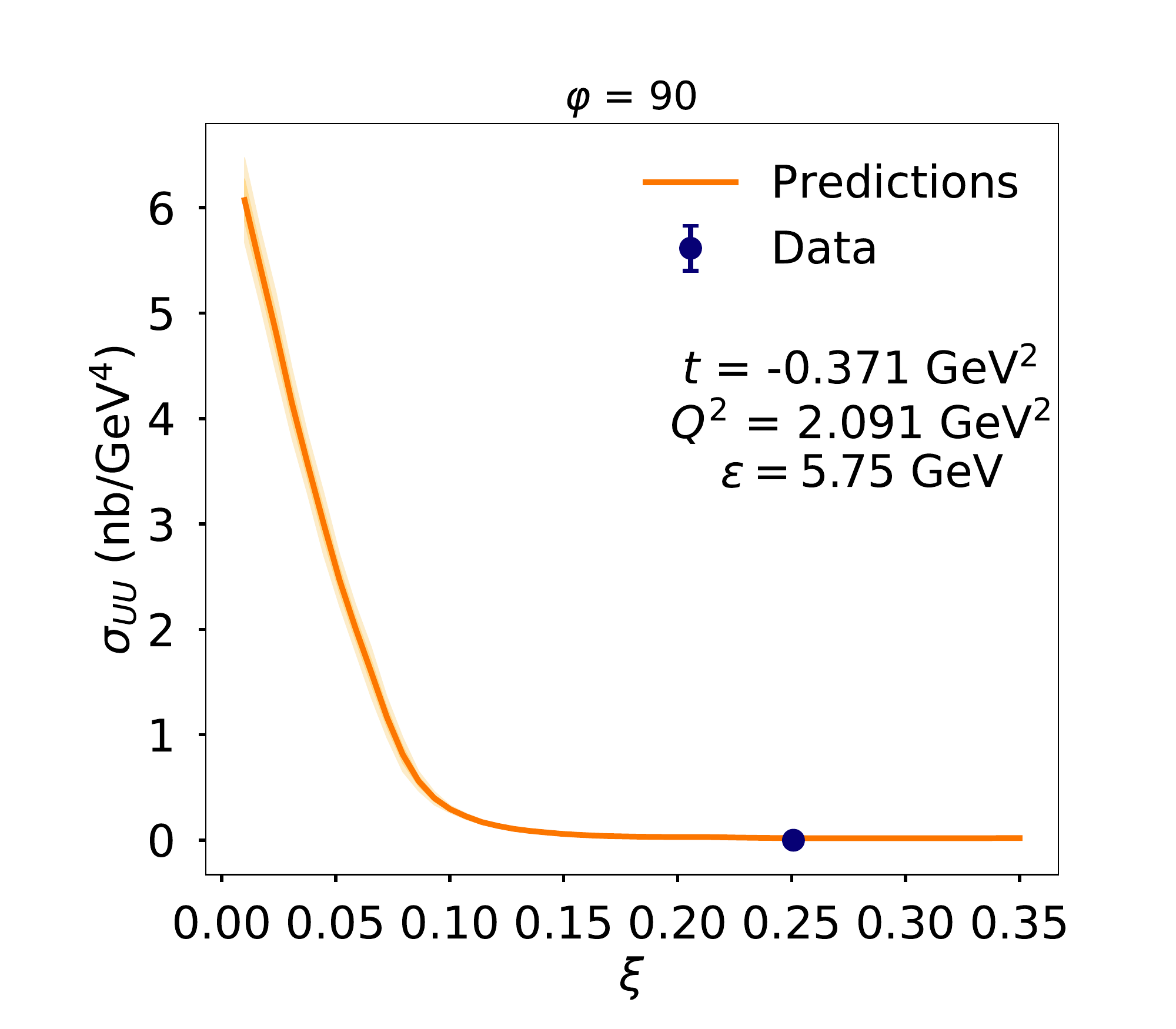} \\
\caption{Predictions across a range of $\xi$ values. $\xi$ is converted to $x_{Bj}$ before being sent to the model. The blue dots in the figure represent the experimental data. The uncertainty in the curves was discussed in Section \ref{uncertainty-explanation}. Notice that for $\xi<0.1$ the cross section is evaluated in a kinematically unphysical region where the neural network can still make predictions.}
\label{fig:xi_dep}
\end{figure}

We aren't restricted to kinematic bins that have already been measured - the model learns to extrapolate between existing bins. We can sweep through regions where data does not exist, and use the model's uncertainty estimate to advise our use of its predictions. In Figure \ref{fig:xi_dep} we illustrate this point for two particular kinematic values.  

This result is consistent with what reported in the upper panel of Fig.\ref{fig:heatmap} where the model's uncertainty in the experimental data domain is very low. Notice that the neural network seems to be able to predict the $t$ dependence much more accurately than the $xi$ dependence. This result could mean, on one side, that the $t$ dependence has a simpler and consistent functional form throughout the phase space than the $\xi$ dependence. The latter will, in fact, be affected by perturbative QCD evolution. On the other hand, the kinematic spread of the data for the $t$ variable covers a wider section than for $\xi$: this feature might also influence the outcome. Clearly, the availability of more data in a wider range in $\xi$ for fixed $t$ will help interpreting this result.  

 While the dataset contains thousands of datapoints, it contains significantly fewer unique kinematic bins; most of the individual points come from kinematic sweeps in the azimuthal angle $\varphi$. The impact of this can be seen empirically in the model's uncertainty estimates. The uncertainty grows quickly as we move between different kinematic regions, while it remained relatively constant between different $\varphi$ values in a fixed region.


\section{Conclusions and outlook}
\label{sec:5}
In summary, a deep learning neural network analysis is essential to extract information from deeply virtual exclusive scattering data in a model independent way.
These reactions are characterized by final states with a large number of kinematic variables and observables, and a consequent growth in complexity of the mapping between the initial variables into the output states.

Our analysis is a necessary first step in a future global extraction of Compton form factors from data. The extraction of information from the data requires an understanding of the features that directly impact such an analysis. 
 In our case, the significant contribution of noise and the relatively small kinematics spread of the data made predicting the cross section a difficult problem even using a deep neural network. Nevertheless, as shown in Table \ref{tab:baselines_results}, an analysis using a DNN was proven the most efficient and accurate method, as compared to other baselines, for extracting information. 
Several additional challenges in our analysis stem from the current lack of an extensive centralized experimental database for exclusive scattering data. 
Another difficulty with this data set is the mismatch in the quantity of data between the  $\varphi$ dependence and the rest of the kinematic variables ($t$, $Q^2$, and $x_{Bj}$). This  results in  the neural network was not  generalizing well across kinematic space. Such challenges can be curtailed through the establishment of a common data set with bench-marking requirements for the publishing of data. This would optimize the usage of data from different experiments in global analyses. 

Our results from FemtoNet suggest many interesting characteristics of DVCS data. In particular, we have shown that there is a substantial and consistent signal in the unpolarized cross section. This is at variance with 
the polarized cross section where the median absolute percentage error is much larger (see Table \ref{tab:baselines_results}).
Furthermore, 
our results extrapolate to kinematics outside of the range covered in  experiment: the DNN seems to be able to generalize in the kinematic variable $t$ more precisely and for a wider range than for the other variables $\xi$ and $Q^{2}$. This suggests, one side, interesting implications for Fourier transforms for 3D visualization from DVCS data. 
On the other hand, the model's apparent inability to extrapolate in $\xi$ could imply an inconsistent function for $\xi$ across the kinematic phase space. 

We also studied the question of how to use experimental errors in neural networks by incorporating them in the error weighted Huber loss (EW Huber, Eq.(\ref{eq:ewhuber})), and also in our accuracy metric (Fig.\ref{fig:accuracy}). 
For this purpose we used pseudo data.
With pseudo data we have control over the error bars and can establish relationships between experimental errors and the uncertainty on the extracted Compton form factors. 

In future studies we will, on one side, extend our approach to analyze other exclusive experiments at Jlab@12 GeV, as well as at the Electron Ion Collider (EIC), while also addressing remaining open questions. We plan to further study both the accuracy and the uncertainty obtained from the DNN analysis of DVCS data. In particular, we will analyze 
the contribution of the background Bethe-Heitler process in the unpolarized cross section that seems to dominate the accuracy of the predictions with respect to the polarized electron scattering. The framework provided in this paper to treat the experimental error in ML affords us the ability to next study the effect of the data errors on the uncertainty of the extracted Compton form factors. 
 This will, in turn, allow us to optimize both the precision and the kinematic domain for experimental measurements  at future colliders needed to efficiently disentangle the observables from the data. 

Finally, our method is flexible to be extended to incorporate lattice QCD results in the data analysis. The latter will play an even more important role in future exclusive experiments analyses than in deep inelastic scattering, towards the ultimate extraction of GPDs from data. 

\acknowledgements
This work was funded by DOE grant DE-SC0016286 and in part by the DOE Topical Collaboration on TMDs (B.K. and S.L.), DOE grant DE-FG03-95ER40965 (M.B.) and SURA grant C2019-FEMT-002-04 and C2020-FEMT-006-05. We gratefully acknowledge the University of Virginia School of Data Science for logistical support; technical assistance by Bryan Wright and Rick Marshall (UVA, Physics), and many insightful physics discussions with Gordon Cates.

\bibliography{BIB}
\end{document}